\documentclass{aa}
\usepackage{graphicx}
\begin{document}
   \title{Compton reflection and iron fluorescence in BeppoSAX observations
of Seyfert type 1 galaxies}

\authorrunning{G. C. Perola et al.}
\titlerunning{Compton reflection and iron fluorescence in Seyfert 1 galaxies}

   \author{G. C. Perola
          \inst{1}
    \and G. Matt
          \inst{1}
 \and M. Cappi
        \inst{2}
 \and F. Fiore
        \inst{3}
 \and M. Guainazzi
        \inst{4}
 \and L. Maraschi
        \inst{5}
 \and P. O. Petrucci
        \inst{5, 6}
 \and L. Piro
         \inst{7}
         }

   \offprints{G. C. Perola, perola@fis.uniroma3.it}

   \institute{Dipartimento di Fisica, Universit\`a degli Studi ``Roma Tre",
              Via della Vasca Navale 84, I--00146 Roma, Italy
 \and Istituto di Astrofisica Spaziale e Fisica Cosmica, CNR, Via Gobetti 101, 
        I--40129 Bologna, Italy   
\and Osservatorio Astronomico di Roma, Via dell'Osservatorio,
        I--00044 Monteporzio Catone, Italy
\and XMM-Newton SOC, VILSPA--ESA, Apartado 50727, E--28080 Madrid, Spain
\and Osservatorio Astronomico di Brera, Via Brera 28, I-20121 Milano, Italy
\and Laboratoire d'Astrophysique de l'Observatoire de Grenoble, B. P. 53X,
      F38041 Grenoble Cedex, France  
\and Istituto di Astrofisica Spaziale e Fisica Cosmica, CNR, Via Fosso del Cavaliere,
               I--00133 Roma, Italy}

   \date{Received; accepted}

\abstract{A sample of nine bright Seyfert 1 and NELG type galaxies, observed
with BeppoSAX, is analyzed to assess on a truly broad band basis (0.1--200 keV)
the issue of the spectral contributions of Compton reflection and iron line
fluorescence from circumnuclear gas.  The empirical description adopted for the 
direct continuum is the commonly used power law with an exponential cut--off.
The most direct test of the theoretical
predictions, namely that the equivalent width of the line, $W_{\alpha}$, and 
the strength $R$ of the
reflection relative to the direct continuum
are closely related to each other, gives a substantially positive result, that is
their mean ratio is very close to expectation, and only a modest spread in the iron
abundance seems implied. The existence of a steep correlation between $R$ and 
the slope $\Gamma$ of the power law is not confirmed. A weak evidence is found
that the existence of a very shallow trend to increase on average with $\Gamma$
cannot be altogether excluded in both $R$ and $W_{\alpha}$, but needs to be tested
with a larger sample. The energy $E_{\rm f}$ in the exponential cut--off spans a range
from about 80 to more than 300 keV. A possible correlation is found, with $E_{\rm f}$
increasing on average with $\Gamma$: if ignored, for instance by keeping $E_{\rm f}$
at a fixed value in a sample study, it could be cause of artificial 
steepening in a correlation between $R$ and $\Gamma$.
}

   \maketitle

\section{Introduction}

Observations with the {\it Ginga} satellite in the 2--18 keV band
of a sample of bright Active Galactic Nuclei (Nandra \& Pounds 1994),
following the discoveries with the same satellite by Piro et al. (1990),
Matsuoka et al. (1990) and Pounds et al. (1990),
definitely confirmed the presence of reprocessed, in addition to the direct,
power law emission, as predicted by e.g.  Guilbert \& Rees (1988), 
Lightman \& White (1988), George \& Fabian (1991), Matt et al. (1991).
The reprocessed radiation consists of two main components. The first is seen
as a broad excess rising above about 10 keV, due to Compton reflection of the power law 
by optically thick, low ionization matter, the second is K$_{\alpha}$ 
fluorescence of mainly ``cold" iron at 6.4 keV. Both appear to be common features in 
Seyfert galaxies of type 1 and intermediate between type 1 and 2 
(objects of the latter type are often referred to as Narrow Emission Line 
Galaxies, NELG, their X-ray spectra are typically affected by a substantial,
order 10$^{22-23}$ H atoms cm$^{-2}$, neutral absorbing column).

Once the inclination angle $i$ between line of sight and material is
adopted, the amplitude of the reflection relative to the power law
can be converted into a measure $R=\Omega/2\pi$ of the solid angle
subtended by the reflector. Through a reanalysis of {\it Ginga}
data for 23 radio--quiet Seyfert 1 and NELG, Zdziarski et al. (1999) 
found evidence of a strong correlation between
$R$ (estimated for a fiducial value of $i=30^{\circ}$) and the photon
slope $\Gamma$ of the power law, which they tentatively describe with
a function where the average 
$R\propto \Gamma^v$, $v$ about 12: $R$ rises from about 0.1
at $\Gamma$=1.7 to about 1 at $\Gamma$=2.1. They argue that the
correlation is not an orientation effect, and, in the framework
of the models where the production of the power law is associated
with a very hot medium that comptonizes soft photons, they tentatively 
attribute it to a feedback of the additional photons from the reflection
on the physical state of that medium. At present, however, its real
significance, at least in its quantitative terms, is in doubt, 
because the existence of a strong correlation between these two variables
in the fit of single observations is likely to bias
substantially the result. Vaughan \& Edelson (2001), for instance,
on the basis of extensive simulations of data obtained with the PCA
instrument onboard the RossiXTE satellite in an energy band similar to that
of the {\it Ginga} data, have shown that a spurious correlation can
be produced. Done \& Nayakshin (2001) proposed yet an alternative possibility,
namely that the correlation could be an artifact of the single--zone ionization
(neutral in practice) model adopted for the reflector: however the simulations of 
{\it Ginga} spectra with their sophisticated model show that the
correlation should be much shallower than the analytical
description quoted above.

The equivalent width $W_{\alpha}$ of the iron fluorescence line, which
inevitably arises in the reflecting material, is expected to be
proportional to $R$. However, a line with $W_{\alpha}$ up to about 100 eV,
that is a substantial fraction of the observed values, in an
optimal geometrical arrangement can be produced also by
marginally thin gas ({\it e.g.} Wo\'zniak et al. 1998). Thanks to the superior 
energy resolution,
the observations with the ASCA satellite of a representative sample of
objects (Nandra et al. 1997) have shown that in most cases the
line is resolved and broader than the velocity spread of the
optical Broad Line Region (BLR). Moreover the first ASCA observation
of MCG--6-30-15 (Tanaka et al. 1995) revealed a profile consistent
with the special and general relativistic effects predicted for
a line originating in the innermost regions of a flat accretion disk
(Fabian et al. 1989). Although the line width in this object turned out
to be rather extreme compared to the others in the Nandra et al. (1997)
sample, it is widely believed (e.g. Fabian et al. 2000) that the
fluorescent gas is most likely associated with the optically thick
accretion disk, and therefore  the broad line $W_{\alpha}$
should correlate with $R$. Unfortunately the ASCA band, being restricted
to less than 10 keV, is inadequate to estimate $R$, and because of the 
variability which typically affects the continuum emission of these objects
(in addition to reported cases of variations in the line, e.g.
Weaver et al. 2001 and references in Fabian et al. 2000), 
the comparison of $W_{\alpha}$, as measured with ASCA, and of $R$ as measured 
with {\it Ginga}, would lack significance.
An alternative approach chosen by Lubi\'nski \& Zdziarski (2001) aims
at looking for a correlation between the broad line
$W_{\alpha}$ and $\Gamma$ in ASCA observations of 25 objects, which should
correspond to the one between $R$ and $\Gamma$. Their Fig. 1
shows indeed a trend, on average, in $W_{\alpha}$ (single gaussian fit) 
to increase with $\Gamma$.
This trend is quantified by fitting a relativistic diskline profile and
an additional narrow gaussian profile to three average spectra,
each one grouped according to the slope: the hardest group
($\Gamma_{\rm av}$=1.62) has broad $W_{\alpha}$=72$^{+17}_{-18}$ eV, 
the softest
group ($\Gamma_{\rm av}$=1.94) has broad $W_{\alpha}$=124$\pm$26, 
with an average
inclination angle (a product of the diskline fit) practically the same
for the three groups, $i\sim45^{\circ}$.
The additional gaussian has an average narrow $W_{\alpha}$ of about 50 eV
in all three groups. The authors interpret this component as originating
in  marginally thin gas: this is a possibility, as mentioned
above, but the alternative one that it might instead be associated with
highly reflecting material well beyond the accretion disk looks at present
at least equally viable. For instance, the case of NGC 4051 observed with 
BeppoSAX in a state of exceptionally low intensity of its power law 
(Guainazzi et al. 1998) shows that the narrow line can be accompanied 
by the right amount of reflection, likely associated with a thick 
molecular torus well beyond the accretion disk,
as predicted (e.g. Ghisellini et al. 1994) in Seyfert type 1 and 2 
unified models (Antonucci 1993). The results in Lubi\'nski \& Zdziarski (2001)
have been recently criticized
by Yaqoob et al. (2002), and should be regarded with some caution.
Besides this criticism, we note that the trend in their Fig. 1, mentioned above
and which is based on individual measurements of  
$W_{\alpha}$ and $\Gamma$, appears to be shallower than the one
between $R$ and $\Gamma$ claimed by Zdziarski et al. (1999).

In this context, the goal of this paper is to provide an assessement
of the observational contribution of the italian--dutch satellite
BeppoSAX (Boella et al. 1997a). This satellite is characterized by: a 
very broad energy band, 0.1--200 keV, covered by its Narrow Field 
Instruments (NFI), namely the LECS (0.1--10 keV, Parmar et al. 1997),
the MECS (1.8--10.5 keV, Boella et al. 1997b) and the PDS (13--200 keV,
Frontera et al. 1997); the energy resolution of the imaging ones (LECS
and MECS)  $\Delta E/E\sim$ 0.08($E$/6~keV)$^{-{1 \over 2}}$, 
which is intermediate between that
of ASCA and of {\it Ginga} at the iron line; the very effective control of the
background in the PDS. These properties have shown their value for the
estimate in several Seyfert 1 galaxies of the spectral parameters of interest 
for this paper, provided their flux in the 2--10 keV
band is at least (2--3)$\times10^{-11}$ erg cm$^{-2}$ s$^{-1}$ 
(e.g. Guainazzi et al 1999a, Perola et al. 1999).

Thanks to the signal detected in the PDS band, the strength of the reflection 
is basically measured in the fit by the excess counts in the 20--50 keV,
where the broad maximum of this component is located, over the power law
whose slope is constrained by the spectral counts in the LECS and MECS.
Determining the slope of the power law is therefore the main issue, and
the confidence beyond the purely statistical errors is related to the
confidence that can be associated, object by object, to the treatment
of additional spectral components, such as the cold and warm (ionized
gas) absorbers and a generic soft excess.  In addition, the power law itself 
may be affected by a high energy ``break'' (as first seen in NGC 4151,
Jourdain et al. 1992), which we shall, as it is often done in the literature,
empirically describe with an exponential factor,
$\exp(-E/E_{\rm f})$. Since the relative strength of the reflection below 
and at its broad maximum
is very sensitive to the Compton recoil in the scattering of higher energy
photons, hence to $E_{\rm f}$, the ability to estimate both $R$ and $E_{\rm f}$ 
thanks to the wide band and 
the efficient background control of the PDS (collimator used in rocking mode)
is an asset of BeppoSAX. Nonetheless caution must be retained on single observation
measurements of the three parameters $\Gamma$, $R$ and $E_{\rm f}$, because
systematic errors on $\Gamma$, arising from an imperfect 
description of the spectrum at low energies,
could be substantial compared to the purely statistical errors.

When the iron line is resolved, it might seem more appropriate to use models,
both for the line and the associated reflection, which incorporates 
Doppler and relativistic effects. This approach would be physically
more self-consistent, and in principle provide in addition the inclination $i$.
However, as confirmed by observations with the {\it Chandra} and XMM-{\it Newton}
satellites (detailed in Sect. 3 for our objects), 
the presence of a substantial narrow contribution is a rather
common property, which is difficult to separate
from the broad contribution with the BeppoSAX resolution, thus making
this approach very tricky, particularly for the objects in our sample,
where the line width is modest, $\sigma\leq0.4$ keV. We have 
therefore chosen to follow a conservative
approach, where the model adopted for the reflection component is
static, and the line is described as a single gaussian. We cared, though,
to verify the outcome of adding to the model a narrow line at 6.4 keV: in general no
improvement in the fit is achieved, moreover, when both a narrow and a broad 
component are found, their combined $W_{\alpha}$ does not differ significantly 
from that obtained with a single gaussian. 

Despite their BeppoSAX observations fulfilled the selection criteria
given in Sect. 2, in order to limit the incidence in our sample of the
systematic errors mentioned above, we prudently excluded objects
with a too complex and variable absorber, like NGC 4151 (Piro et al. 2001),
NGC 3516 (Costantini et al. 2000, Guainazzi et al. 2001) and NGC 2992
(Gilli et al. 2000). The observation of MCG--6-30-15 (Guainazzi
et al. 1999b) confirmed the ASCA finding (Tanaka et al.
1995) of a very broad and asimmetrical iron line, and was also excluded 
because it could not be treated homogeneously with the rest of the sample.

The paper is organized in the following manner. Sect. 2 describes
the sample of the observations used. Sect. 3 describes the spectral
fitting procedures and the results. Sect. 4 illustrates the distribution
of the parameters in connection with the reprocessing. Sect. 5 is devoted 
to discussion and conclusions.

\section{The sample data}

The sample contains observations of objects with flux in the 2--10 keV
band greater than 3$\times10^{-11}$ erg cm$^{-2}$ s$^{-1}$, duration 
longer than 50000 s in the MECS. For some objects, with separate observations 
shorter than 
50000 s, after verifying with a hardness ratio analysis, as well as independent
fits, that no spectral changes had occurred of such entity as to impair
the reliability of the outcome relevant to the present paper, we
merged two (NGC 5506, Mrk 509) or three observations (IC4329A(2)).
Several of these observations are the subject of papers already
published, whose reference will be given in Sect. 3.

\begin{table*}
\begin{center}
\caption{The sample}
\scriptsize
\begin{tabular}{||c|c|c|c|c|c|c|c|c|c||}
\hline
\hline
~& ~& ~& ~ & ~ & ~ & ~ & ~ & ~ & ~\cr
Source & Start  & LECS & MECS  & PDS & $z^a$ & N$_{\rm g}^b$ & $F_{\rm o}^c$ 
& $F_{\rm na}^c$ & log$L_{\rm na}^d$ \cr
name & date & T$_{\rm exp}$~(s) & T$_{\rm exp}$~(s) & T$_{\rm exp}$~(s) & & 
($\times10^{20}$ cm${-2}$) & (2--10~keV) & (2--10~keV) & (2--10~keV) \cr
~& ~& ~& ~ & ~ & ~ & ~ & ~ & ~ & ~\cr
\hline
~& ~& ~& ~ & ~ & ~ & ~ & ~ & ~ & ~\cr
MCG~8-11-11 &  1998-Dec-07 & 29840 & 74865 & 72472 & 0.020 & 20.3$^1$ & 5.61 &
5.73 & 43.62  \cr
~& ~& ~& ~ & ~ & ~ & ~ & ~ & ~ & ~\cr
MCG--5-23-16 &  1998-Apr-24 & 35825 & 76941 & 65048 & 0.008 & 8.4$^2$ & 8.73 &
10.00 & 43.06 \cr
~& ~& ~& ~ & ~ & ~ & ~ & ~ & ~ & ~\cr
NGC~3783 &  1998-Jun-06 & 69135 & 153870 & 140814 & 0.009 & 9.6$^2$ & 6.44  & 
6.86 & 43.00 \cr
~& ~& ~& ~ & ~ & ~ & ~ & ~ & ~ & ~\cr
NGC~4593 &  1997-Dec-31 & 31426 & 84057 & 76934 & 0.009 & 2.0$^1$ & 3.72 & 
3.79 & 42.74 \cr
~& ~& ~& ~ & ~ & ~ & ~ & ~ & ~ & ~\cr
IC~4329A~(1) &  1998-Jan-02 & 25097 & 81826 & 75214 & 0.016 & 4.5$^1$ & 
13.15 & 13.83 & 43.81 \cr
~& ~& ~& ~ & ~ & ~ & ~ & ~ & ~ & ~\cr
~& 1998-Jul-17 & ~& ~ & ~ & ~ & ~ & ~ & ~ & ~\cr
IC~4329A~(2)  &  1998-Jul-21 & 38245 & 106140 & 94772 & & & 10.68 & 11.18 &
43.71 \cr
~& 1998-Jul-27 & ~& ~ & ~ & ~ & ~ & ~ & ~ & ~\cr
~& ~& ~& ~ & ~ & ~ & ~ & ~ & ~ & ~\cr
~& 1997-Jan-30 & ~& ~ & ~ & ~ & ~ & ~ & ~ & ~\cr
NGC~5506 & ~ & 25101 & 77491 & 70842 & 0.006 & 4.3$^2$ & 7.20 & 9.57 &
42.79 \cr
~& 1998-Jan-14 & ~& ~ & ~ & ~ & ~ & ~ & ~ & ~\cr
~& ~& ~& ~ & ~ & ~ & ~ & ~ & ~ & ~\cr
NGC~5548 & 1997-Aug-14 & 85444 & 312810 & 239320 & 0.017 & 1.6$^2$ & 3.55 &
3.61 & 43.27 \cr
~& ~& ~& ~ & ~ & ~ & ~ & ~ & ~ & ~\cr
~& 1998-May-18 & ~& ~ & ~ & ~ & ~ & ~ & ~ & ~\cr
Mrk~509 & ~& 41786 & 87834 & 82102 & 0.035 & 4.4$^2$ & 5.66 & 5.69 & 44.10 \cr
~& 1998-Oct-11 & ~& ~ & ~ & ~ & ~ & ~ & ~ & ~\cr
~& ~& ~& ~ & ~ & ~ & ~ & ~ & ~ & ~\cr
NGC~7469 & 1999-Nov-23 & 99114 & 249780 & 243020 & 0.017 & 4.8$^1$ &  3.73 &
3.76 & 43.29 \cr
~& ~& ~& ~ & ~ & ~ & ~ & ~ & ~ & ~\cr
\hline
\hline
\end{tabular}
\end{center}
$^a$ Redshift values from V\'eron-Cetty \& V\'eron (1993) \par
$^b$ Galactic H column from: (1) Elvis et al. (1989) and
(2) Murphy et al. (1996) \par
$^c$ $F_{\rm o}$: Flux as observed; $F_{\rm na}$:  corrected for absorption; 
both in units of 10$^{-11}$ erg cm$^{-2}$ s$^{-1}$ \par
$^d$ Log of luminosity corrected for absorption in erg  s$^{-1}$ 
(H$_o$=75 km s$^{-1}$ Mpc$^{-1}$)
\end{table*}

Table 1 lists for each object: the observation start-dates and the good
exposure times in the LECS, MECS and PDS, the redshift and the galactic
hydrogen column, $N_{\rm g}$, adopted, the flux in the 2--10 keV band as
observed, $F_{\rm o}$, corrected for any absorption as from the model 
fit, $F_{\rm na}$, 
and its corresponding luminosity ($H_o$=75 km s$^{-1}$ Mpc$^{-1}$).

The reduction procedure and screening criteria adopted are standard
(see e.g. Guainazzi et al. 1999a). Concerning the PDS data, the
Variable Rise Time (VRT) option was generally adopted, except for
the observation of NGC 4593, where the Fixed Rise Time (FRT) was used.
The PDS net spectra were obtained from direct subtraction of the off- from 
the on--source products. 
{We verified that for none of the sample objects
there was a significant contamination by hard sources in both the on-
and the off--source field of view.}
Concerning the data from the LECS and the MECS images, the background 
was subtracted using spectra from blank sky files at the source position 
in the detectors.

The exposure times in the LECS are much shorter than in the MECS,
because the former instrument is operated only during the night--time
fraction of each orbit. Since the incident flux is generally recorded
as variable during each observation, to take care of this inhomogeneity
in the time coverage, in the fits the normalization of the LECS to
the MECS was left as a free parameter. The normalization of the
PDS to the MECS was instead held fixed at 0.8 (VRT selection) or
0.86 (FRT selection, Fiore et al. 1999). For the LECS data
we used the version Sept. 1997 of the transfer matrix, and limited
the range to 0.1--4.0 keV, where this matrix is more reliable. 
The energy bins chosen  correspond to about one third
of the instrument resolution, which itself is energy dependent.

\section{Spectral analysis}

The analysis of the spectral counts was performed using the software
package XSPEC (Arnaud 1996, version 11.0.1). The Baseline Model Spectrum
(BMS) includes: a photon power law (PL) with exponential cut--off,
$AE^{-\Gamma}\exp(-E/E_{\rm f})$, 
together with a reflection component (RC) from
a cold slab isotropically illuminated by the PL photons, 
subtending the solid angle 
$R=\Omega/2\pi$ and with inclination cos~$i$ to the line of sight
(Magdziarz \& Zdziarski 1995, module {\sc pexrav});
a uniform cold absorber column at the source, $N_{\rm s}$, in addition to
$N_{\rm g}$ given in Table 1; a uniform warm absorber at the source,
as described in the module {\sc absori}, with column $N_{\rm w}$ and 
ionization
parameter $X_{\rm i}=L_{\rm i}/nD^2$ erg cm s$^{-1}$, where $n$ is the 
gas density and 
$D$ its distance from the source of ionizing luminosity $L_{\rm i}$ in the
interval 5 eV to 20 keV (for  $L_{\rm i}$, assumed with a power law spectrum,
we set the slope equal to $\Gamma$; the temperature of the absorbing
gas was fixed as a rule at 3$\times10^4$ K); a gaussian line to represent
the iron K$_{\alpha}$ fluorescence, with energy $E_{\alpha}$, width 
$\sigma_{\alpha}$,
intensity $I_{\alpha}$, accompanied by a gaussian of identical width, to
represent the iron K$_{\beta}$ (with $E_{\beta}$ and $I_{\beta}$ set equal
to 1.1$E_{\alpha}$ and 0.11$I_{\alpha}$, George \& Fabian 1991).
For both reflection 
and absorption we adopted the element abundances in Anders \& Grevesse (1993).
All fits were performed with cos~$i$ fixed equal to 0.9 ($i=26^{\circ}$).
Alternatively cos~$i$ was left free with $R$ fixed equal to 1. Because the 
dependence on cos~$i$ of the RC shape is modest, all other parameters turned
out identical in practice to those obtained with the angle fixed, and
the differences in $\chi^2$ are insignificant.

\begin{table*}
\begin{center}
\caption{Results of the spectral analysis}
\scriptsize
\begin{tabular}{||c|c|c|c|c|c|c|c|c|c|c||}
\hline
\hline
~& ~& ~& ~ & ~ & ~ & ~ & ~ & ~ & ~ & ~\cr
Source & $\Gamma$ & $E_{\rm f}$ & $R$ & $E_{\alpha}$ & $\sigma_{\alpha}$ & 
$I_{\alpha}^a$ & $W_{\alpha}$ & $\chi^2$/dof & cos~$i$ & $\chi^2$/dof \cr
name & & (keV) & (cos~$i$=0.9) & (keV) & (keV) & & (eV) & & ($R$=1) & \cr
~& ~& ~& ~& ~ & ~ & ~ & ~ & ~ & ~ & ~\cr
\hline
~& ~& ~& ~& ~ & ~ & ~ & ~ & ~ & ~ & ~\cr
MCG~8-11-11 & 1.85$^{+0.09}_{-0.05}$ & 166$^{+215}_{-74}$ & 
1.09$^{+0.80}_{-0.39}$ & 6.49$\pm$0.14 & $<0.37$ & 6.9$^{+3.7}_{-2.4}$ &
117$^{+63}_{-40}$ & 132.3/141 & 0.96$^{+0.04}_{-0.49}$ & 132.6/141 \cr
~& ~& ~& ~& ~ & ~ & ~ & ~ & ~ & ~ & ~\cr
MCG--5-23-16 & 1.81$\pm$0.05 & 147$^{+70}_{-40}$ & 0.66$^{+0.25}_{-0.20}$ & 
6.43$\pm$0.10 & $<0.28$  & 10.0$^{+3.9}_{-3.0}$ &
96$^{+37}_{-28}$ & 139.5/135 & 0.51$^{+0.24}_{-0.17}$ & 140.7/135 \cr
~ & ~& ~& ~& ~ & ~ & ~ & ~ & ~ & ~ & ~\cr
NGC~3783 &  1.77$\pm$0.04 & 156$^{+37}_{-40}$ & 0.63$^{+0.20}_{-0.17}$ & 
6.38$\pm$0.09 & 0.27$^{+0.15}_{-0.13}$  & 11.7$^{+2.8}_{-2.7}$ &
161$^{+39}_{-37}$ & 150.4/148 & 0.46$^{+0.19}_{-0.14}$ & 148.0/148 \cr
~& ~& ~& ~& ~ & ~ & ~ & ~ & ~ & ~ & ~\cr
NGC~4593 &  1.94$^{+0.06}_{-0.05}$ & $>222$ & 1.26$^{+0.65}_{-0.40}$ & 
6.41$\pm$0.16 & 0.34$^{+0.36}_{-0.21}$  & 9.1$^{+4.4}_{-2.8}$ &
231$^{+111}_{-72}$ & 148.2/146 & $>0.63$ & 149.0/146 \cr
~ & ~& ~& ~& ~ & ~ & ~ & ~ & ~ & ~ & ~\cr
IC~4329A~(1) &  1.89$\pm$0.04 & 325$^{+277}_{-105}$ & 0.63$^{+0.17}_{-0.14}$ & 
6.43$\pm$0.17 & 0.30$^{+0.30}_{-0.26}$  & 14.5$^{+5.7}_{-5.6}$ &
104$^{+41}_{-40}$ & 146.9/141 & 0.49$^{+0.14}_{-0.12}$ & 148.2/141 \cr
~ & ~& ~& ~& ~ & ~ & ~ & ~ & ~ & ~ & ~\cr
IC~4329A~(2) &  1.90$\pm$0.05 & 262$^{+204}_{-84}$ & 0.73$^{+0.22}_{-0.18}$ & 
6.54$\pm$0.14 & 0.31$^{+0.25}_{-0.15}$  & 13.5$^{+5.1}_{-4.0}$ &
123$^{+46}_{-36}$ & 154.4/140 & 0.58$^{+0.19}_{-0.14}$ & 153.9/140 \cr
~ & ~& ~& ~& ~ & ~ & ~ & ~ & ~ & ~ & ~\cr
NGC~5506 &  2.02$^{+0.09}_{-0.08}$ & $>298$ & 1.20$^{+0.45}_{-0.35}$ & 
6.52$^{+0.10}_{-0.09}$ & 0.21$^{+0.15}_{-0.10}$  & 14.2$^{+4.2}_{-3.2}$ &
152$^{+45}_{-34}$ & 85.3/82 & $>0.75$ & 85.9/82 \cr
~ & ~& ~& ~& ~ & ~ & ~ & ~ & ~ & ~ & ~\cr
NGC~5548 &  1.62$^{+0.04}_{-0.05}$ & 147$^{+64}_{-33}$ & 
0.54$^{+0.20}_{-0.13}$ & 
6.41$\pm$0.07 & $<0.13$  & 4.3$^{+1.1}_{-0.9}$ &
106$^{+26}_{-23}$ & 175.1/148 & 0.37$^{+0.21}_{-0.12}$ & 176.7/148 \cr
~ & ~& ~& ~& ~ & ~ & ~ & ~ & ~ & ~ & ~\cr
Mrk~509 & 1.58$^{+0.09}_{-0.08}$ & 67$^{+30}_{-20}$ & 0.58$^{+0.39}_{-0.30}$ & 
6.64$^{+0.34}_{-0.24}$ & $<0.87$  & 5.5$^{+4.5}_{-2.8}$ &
93$^{+76}_{-46}$ & 135.7/132 & 0.45$^{+0.50}_{-0.33}$ & 137.3/132 \cr
~ & ~& ~& ~& ~ & ~ & ~ & ~ & ~ & ~ & ~\cr
NGC~7469 &  1.88$^{+0.05}_{-0.07}$ & 164$^{+196}_{-65}$ & 
0.50$^{+0.29}_{-0.25}$ & 
6.44$^{+0.12}_{-0.11}$ & $<0.47$  & 5.2$^{+1.9}_{-1.7}$ &
139$^{+51}_{-44}$ & 168.4/148 & 0.35$^{+0.31}_{-0.20}$ & 168.0/148 \cr
~ & ~& ~& ~& ~ & ~ & ~ & ~ & ~ & ~ & ~\cr
\hline
\hline
\end{tabular}
\end{center}
$^a$ Line intensity in units of 10$^{-5}$ photons cm$^{-2}$ s$^{-1}$ 
\end{table*}

The results on the parameters of main interest for this paper are
collected in Table 2. The line parameters are given in the source
frame, and the line intensity $I_{\alpha}$ is also expressed as an
equivalent width, $W_{\alpha}$. The errors represent the 90\% confidence
interval for two parameters of interest ($\Delta\chi^2$=4.61, Lampton
et al. 1976),
a realistic choice dictated by the statistical correlation which
exists between couples such as ($\Gamma$, $R$), ($\Gamma$, $E_{\rm f}$), 
($\sigma_{\alpha}$, $I_{\alpha}$).

To be noted that in general the value of $E_{\alpha}$ is perfectly 
consistent with
6.4 keV, the value for ``cold" iron (in agreement with previous findings,
e.g. Nandra \& Pounds 1994, Nandra et al. 1997, Lubi\'nski \& Zdziarski 2001).
In two objects (NGC 5506 and Mrk 509) the (per se marginal) evidence
of a larger value finds explanation in recent XMM-{\it Newton} 
results, reported below.

Next we briefly describe the case of each source, including the
best fit values of the other BMS parameters and additional spectral
components adopted to improve the quality of the fit
when the BMS gave an unacceptable $\chi^2$. Generally  $X_{\rm i}$
is very poorly constrained, hence we give only its best fit value.
Reference is also given to some recent results on the iron line.

{\it MCG~8-11-11}. This observation was published by Perola et al. (2000),
to which we refer for the choice of neglecting in the fit the LECS
counts below 0.4 keV. $N_{\rm s}$ is negligible compared to $N_{\rm g}$, 
the warm
absorber has $N_{\rm w}$=(6$^{+9}_{-6}$)$\times10^{20}$ 
cm$^{-2}$ with $X_{\rm i}\sim$7
erg cm s$^{-1}$. The iron line is not resolved, the upper limit 
on its width is comparable
to that  of the lines resolved in our sample.

{\it MCG--5-23-16}. The cold absorber in this NELG is substantial,
$N_{\rm s}$=(1.6$\pm$0.9)$\times10^{22}$ cm$^{-2}$; no 
signature is detected of a warm
absorber. In ASCA (Weaver et al. 1997) and RossiXTE (Weaver et al.
1998) observations (comparable in flux to this one) a strong 
($W_{\alpha}\sim$250 eV) and broad iron line was consistently found. In
the BeppoSAX  observation the line is 
much narrower than $\sigma_{\alpha}\sim$ 0.4 keV, the value  obtained from a 
single gaussian description in both the ASCA and the RossiXTE data sets
(Weaver et al. 1998); the line intensity is about twice that of the narrow
gaussian, but about one third of the sum of the narrow and broad
gaussians, as measured by Weaver et al. (1997) in their fit to the ASCA 
data. While this is probably yet another case of variations in the iron
line (see also Smith \& Done 1996 for a formally significant difference
in $W_{\alpha}$ between two {\it Ginga} observations made a few days apart),
we note that, contrary to expectation, the amount of reflection
measured with RossiXTE (for a choice of $E_{\rm f}$=200 keV, the closest
to our best fit value of the three adopted; $\Gamma$ is identical to
our best fit value; Weaver et al. 1998) is definitely not larger
than in the BeppoSAX observation.  Finally we note that a {\it Chandra}
HETGS observation, briefly described in Weaver (2001), shows a narrow
(FWHM less than 3000 Km s$^{-1}$) line with $W_{\alpha}\sim$ 90 eV, similar
to our value.

{\it NGC 3783}. The BMS fit is poor ($\chi^2$=208/150), with an excess in
the residuals at low energies. With the addition of a line the
$\chi^2$ reduces to 160/147, with $E_{\rm l}$=0.59 keV and $W_{\rm l}$=140 eV:
 its strength is however almost ten times larger than that of the 
OVII emission lines detected by Kaspi et al. (2001)
in a {\it Chandra} HETGS observation (when the source was fainter by
only about 15\%). A formally better improvement is achieved if, instead of a line,
a black body is added, provided the temperature of the warm
absorber is set equal to 10$^5$ K: the parameters given in Table 2
refer to this fit. $N_{\rm s}$ is negligible compared to $N_{\rm g}$,
$N_{\rm w}$=(9.6$^{+3.2}_{-1.4}$)$\times10^{21}$ 
cm$^{-2}$, with $X_{\rm i}\sim$15 erg cm s$^{-1}$,
the black body $kT$=0.18$^{+0.02}_{-0.03}$ keV. The iron line is definitely
resolved. The {\it Chandra} HETGS observation (Kaspi et al. 2001)
has revealed a narrow 
line at 6.4 keV, with an upper limit on the width ($\sigma$ less than
0.030 keV) that places its origin beyond the BLR:
its intensity of (6.6$\pm$2.1)$\times10^{-5}$ cm$^{-2}$ s$^{-1}$ is about 
55\% of that measured 
by us. If we include in the fit a narrow, in addition to the broad line,
no improvement is attained in $\chi^2$ and the intensity of the narrow 
component is not well constrained, such that a value identical
to that from the {\it Chandra} observation is acceptable.  In a
recent publication (De Rosa et al. 2002) on this BeppoSAX observation, 
the same model adopted by us (their Model E) yields somewhat different
values, in particular $\Gamma$=1.86$\pm$0.03, $R$=0.71$^{+0.20}_{-0.28}$, 
$W_{\alpha}$=210$\pm$45, together with
$N_{\rm w}$ about twice larger. The discrepancy is due to the different code 
adopted to describe the warm absorber, and it is not such as to
influence the conclusions of the analysis following in Sect. 4. We note on the other 
hand the good agreement in our values of $\Gamma$ and $N_{\rm w}$ with those obtained
by Kaspi et al. (2001) in fitting the ``line free zones'' of the HETGS spectral
data with a power law and a warm absorber.

{\it NGC 4593}. This observation was published by Guainazzi et al. (1999a).
$N_{\rm s}$ is negligible compared to $N_{\rm g}$, 
and $N_{\rm w}$=(2.3$^{+1.1}_{-0.9}$)$\times10^{21}$ cm$^{-2}$
with $X_{\rm i}\sim$9 erg cm s$^{-1}$. The iron line is resolved, $W_{\alpha}$
together with $R$ are the largest in our sample.

{\it IC 4329A}. Observation (1) was published by Perola et al. (1999).
The average flux of observation (2) is 20\% lower, yet there are no 
statistically significant differences between the two sets of parameters
 (the same holds for the three pointings merged, when fitted individually,
despite a 25\% difference between the lowest and the highest of their
flux levels).
The mean values of the absorbers are: 
$N_{\rm s}$=(2.7$\pm$1.0)$\times10^{21}$ cm$^{-2}$, 
$N_{\rm w}$=(2.8$^{+1.2}_{-1.0}$)$\times10^{21}$ cm$^{-2}$ with 
$X_{\rm i}\sim$5 erg cm s$^{-1}$.
The iron line is resolved. Given the strength of the source (the
brightest of the sample) and the correspondingly good statistics,
Perola et al. (1999) attempted a fit with a narrow gaussian plus
a relativistic disk profile, and obtained marginal evidence for the
existence of the narrow component. Another marginal detection of the
narrow component in a simultaneous ASCA and RossiXTE observation
is presented by Done et al. (2000).  In a XMM-{\it Newton} 
observation with the EPIC instrument (Gondoin et al. 2001), at a flux level 25\% higher
than in observation (1), the narrow ($\sigma$ less than 0.06 keV)
line is detected, with $W_{\alpha}$=43$\pm$1 eV; a broad component is seemingly
absent, but unfortunately the authors do not provide an upper limit 
on its equivalent width. A simultaneous and relatively short
(34000 s in the MECS) BeppoSAX observation is used, together with
the EPIC data, by Gondoin et al. (2001) 
to estimate $\Gamma$=1.93$\pm$0.03 and $R$=1.1$\pm$0.3: we believe that the 
estimate of $R$ is incorrect, because obtained with a PDS to MECS normalization 
factor equal to an improbable value of 0.7; by fitting the BeppoSAX data,
as retrieved from the public archive, with this
factor equal to 0.86 (for FRT selection, see Sect. 2), we obtain best fit
values of $R$=0.54 and $\Gamma$=1.91, fully consistent with those in
Table 2.

{\it NGC 5506}.  The two observations here merged have an almost identical
average flux level, and when fitted individually yield similar values
of the spectral parameters.  
The BMS fit with all three instruments gives an unacceptable
$\chi^2$=180/117, due to an excess in the residuals well below the strong
photoelectric cut--off. Since the parameters of interest do not
crucially depend on how this excess is modeled, we have chosen to
present in Table 2 the results obtained excluding the LECS. This NELG
has a substantial $N_{\rm s}$=(3.7$\pm$0.2)$\times10^{22}$ 
cm$^{-2}$, and no signatures of a 
warm absorber. The iron line is resolved and $E_{\alpha}$ is marginally
inconsistent with 6.4 keV. This line is likely a 
blend of the two components recently revealed in a simultaneous
XMM-{\it Newton} (with EPIC) and BeppoSAX observation (flux level similar to ours,
Matt et al. 2001), one unresolved at 6.4 keV, the other resolved
($\sigma$ about 0.25 keV) at 6.75 keV, the two with comparable intensities 
whose sum is consistent with the one in our observation. 
Matt et al. (2001) present arguments, centred on the interpretation
of the iron edge seen at 7.1 keV, that a substantial fraction of the
RC in this object is very likely associated with the narrow line
component. Evidence supporting this interpretation can be found in
a spectral variability investigation conducted with RossiXTE by
Lamer et al. (2000).

{\it NGC 5548}. This observation was published by Nicastro et al. (2000).
They find internally evidence of a correlation between $\Gamma$ and
the flux (previously discovered in {\it Ginga} observations by Magdziarz
et al. 1998, confirmed with RossiXTE observations by Chiang et al.
2000), which has some impact on the quality of our fit.
In a {\it Chandra} LETGS observation Kaastra et al. (2002)
find evidence of a very complex warm absorber, of a soft excess
and of low energy emission lines. Following Kaastra \& Barr (1989),
who first reported a soft excess in EXOSAT observations, they model
it as a ``modified black body'' with temperature $kT\sim$0.1 keV.
They suggest that the NVII emission line at 0.5 keV could be identified
with the low energy line detected by Nicastro et al. (2000) in the
BeppoSAX observation. We therefore added to the BMS a soft component,
as described by the module DISKBB in XSPEC, with $kT$ fixed at 0.1 keV,
and a line with $E_{\rm l}$ fixed at 0.5 keV. The fit yields 
$W_{\rm l}$=76$^{+61}_{-54}$ eV, compatible with the {\it Chandra} result.
$N_{\rm s}$ turns out negligible compared to 
$N_{\rm g}$, while
$N_{\rm w}$=(2.4$\pm$0.7)$\times10^{21}$ cm$^{-2}$ (with 
$X_{\rm i}\sim$14 erg cm s$^{-1}$) is marginally consistent with the column
density of the dominant component in the warm absorber as estimated
by Kaastra et al. (2002).
The iron line is unresolved and significantly narrower (less than
0.13 keV) than in any of the other sample objects. A {\it Chandra} HETGS
observation (Yaqoob et al. 2001), with the source at a flux level 
about 30\% lower, has revealed a narrow line at 6.4 keV 
with $\sigma_{\alpha}$=41$^{+32}_{-24}$ eV, corresponding to a velocity spread
consistent with an origin in the outer BLR (a substantial contribution
to the line from thick matter beyond that region is not excluded). The
intensity of this line matches the one measured by us. Using the
same prescriptions for a relativistic disk line used by Yaqoob et al.
(2001) to obtain an upper limit of 7$\times10^{-5}$ cm$^{-2}$ s$^{-1}$, 
we find a more stringent limit of 4$\times10^{-5}$ cm$^{-2}$ s$^{-1}$: this 
is a factor about 2 less than in earlier ASCA records of a substantially broader 
line, as reported in Yaqoob et al. (2001) or in Chiang et al. (2000). 
In a spectral variabilty study with RossiXTE, Chiang et al. (2000)
found the intriguingly paradoxical (for the reprocessing
paradigm) result that $W_{\alpha}$ is anticorrelated to $R$.

{\it Mrk 509}.  The two observations and the results from their
merging were published by Perola et al. (2000): in that paper it is
shown that the BMS fit leaves the line width practically
unconstrained (about 3 keV) and that in this respect the situation changes
radically if a soft component is added in the form of a power law 
(preferred to a black body). This component 
has $\Gamma_{\rm s}$=2.5$^{+0.9}_{-0.2}$ and normalization 
1.1$\times10^{-2}$ cm$^{-2}$ s$^{-1}$ keV$^{-1}$
at 1 keV. $N_{\rm s}$ is negligible compared to $N_{\rm g}$, and 
$N_{\rm w}$=(4$^{+8}_{-2}$)$\times10^{20}$ cm$^{-2}$
with $X_{\rm i}\sim$16 erg cm s$^{-1}$. The line width is poorly constrained
and the line energy is marginally
inconsistent with 6.4 keV. A XMM-{\it Newton} observation (with EPIC, Pounds et al. 
2001) has revealed the presence of a resolved cold and of a broader ionized
(6.9 keV) components. The $W$ of the former is about twice that of the latter,
their sum is consistent with the value measured by us. The source,
however, was almost 3 times fainter than in our observation, thus the
blend must have experienced a change in intensity by a comparable factor.

{\it NGC 7469}. The BMS fit gives a marginally acceptable $\chi^2$=191/151,
but, like in the case of Mrk 509 as illustrated in Perola et al. (2000), 
the line
width is practically unconstrained (about 2 keV). The addition of a
soft excess in the form of a power law changes the
situation radically. The slope of the soft power law is
 $\Gamma_{\rm s}$=3.3$\pm$0.5, its normalization 
5.5$\times10^{-3}$ cm$^{-2}$ s$^{-1}$  keV$^{-1}$ at 1 keV.
$N_{\rm s}$ is negligible compared to $N_{\rm g}$, and 
$N_{\rm w}$=(6.7$^{+4.3}_{-3.7}$)$\times10^{20}$ cm$^{-2}$
with $X_{\rm i}\sim$0.13 erg cm s$^{-1}$. The line is
not resolved, the upper limit
on its width is comparable to that of the lines resolved in our sample.
In a spectral variabilty study with 
RossiXTE, Nandra et al. (2000) detected changes in the line intensity
which are not correlated with the absolute normalization of the 
reflection component, but they themselves doubt the reliability of
their estimate of the second of the two parameters.
 We note that George et al. (1998) did not find evidence of a soft
excess in their analysis (above 0.6 keV) of a ASCA observation
with a flux 25\% lower than in our data: thus the additional soft
component, that we suggest here to provide a better description of
the continuum of this observation, should be variable. We further note
that the value of $\Gamma$ in Table 2 is in good agreement with
$\Gamma$=1.92$\pm$0.02 obtained by Nandra et al. (2000) from the
integrated fit of the data in the study mentioned above, 
with an average flux only 10\% lower than in our observation.

\section{An analysis of the parameters}

This Section is devoted to illustrate the distribution of the
parameters in the sample, in relation with the reprocessing of the PL.

\subsection{$W_{\alpha}$ and R}

Irrespective of model details, under the hypothesis that the
line emission is entirely associated with optically thick material,
to a good approximation $W_{\alpha}$ should correlate linearly with
the value of $R$ obtained for an arbitrarily fixed inclination angle.
Apart from the one introduced by the measurements errors, an 
intrinsic scatter in $W_{\alpha}$ about the correlation can be caused
by a variance in the iron abundance (for differences within
a factor 2 of the cosmic value, the dependence of $W_{\alpha}$ on the
iron abundance is approximately linear, see Matt et al. 1997). 
Furthermore, for a given $R$, $W_{\alpha}$ may differ according to the
value of $\Gamma$, because it is proportional to the ratio of the photon
number above 7.1 keV responsible for the fluorescence to the 
photon flux density at the line energy. The plot in Fig. 1
appears scatter dominated, and because of the rather large
errors on both parameters little can be said about individual
deviations from the hypothetical linear trend.  The Spearman rank--order
(Press et al. 1992) test, which does not take the errors into account,
yields a correlation coefficient which is positive, $r_{\rm s}$=0.43,
but with a large, 0.2, significance level for this quantity deviating from zero.
Following another approach, where the error bars
are also ignored, but which seems more useful in this context, 
we can compute for each object the ratio
of the best fit value of $W_{\alpha}$ to the value expected, given
the $\Gamma$ of the PL and the best fit value of $R$ for cos~$i$=0.9.
We did that by using the $\Gamma$ and angular dependence of $W_{\alpha}$
given in George \& Fabian (1991), rescaled to the Anders \& Grevesse
(1993) abundances (see Matt et al. 1997). We obtain 0.97 for
the mean value, and 0.29 for the standard deviation of the
sample. A similar outcome (mean value 0.88, standard deviation
0.27) obtains when we proceed instead with R fixed equal to 1
and the best fit values of cos~$i$ to estimate the expected $W_{\alpha}$.
In our sample, then, on average the observed $W_{\alpha}$ is of the
right amount expected from the simultaneously measured strength
of the RC relative to the PL for the cosmic value of the element
abundances. The standard deviation, being inevitably affected by the 
large errors on both $W_{\alpha}$ and $R$  can be considered a limit on
the variance of the iron abundance in our sample.

Concerning $R$, we note the concentration (6 out of 9 objects)
of best fit values in the interval 0.50--0.73, or the corresponding 
concentration of cos~$i$ (for $R$=1) in the interval 0.35--0.58. 
The absence of objects with cos~$i$ less than 0.35 (inclined more than 
about 70$^{\circ}$) can be understood in the frame of the Seyfert type 1 and 
type 2 unified models (Antonucci 1993), where
the first type takes the appearence of the second at large inclinations.
The paucity of objects with small, relative to intermediate
inclinations could well be a statistical fluctuation, given the
modest sample size, but it may  reflect intrinsic
differences in $R$, with objects where the intrinsic $R$ is indeed 
substantially less than unity.

\begin{figure}
   \centering
   \includegraphics[width=9.cm]{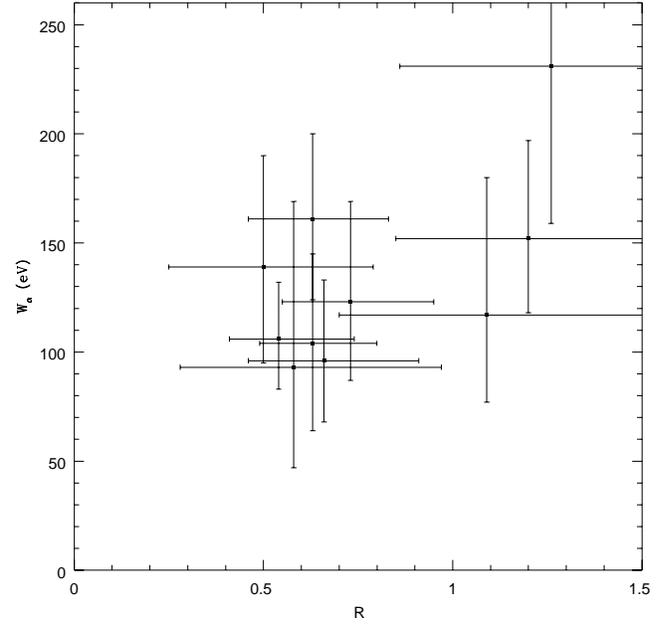}
\caption{$W_{\alpha}$ plotted versus $R$.
}
\end{figure}

\begin{figure}
   \centering
   \includegraphics[width=9.cm]{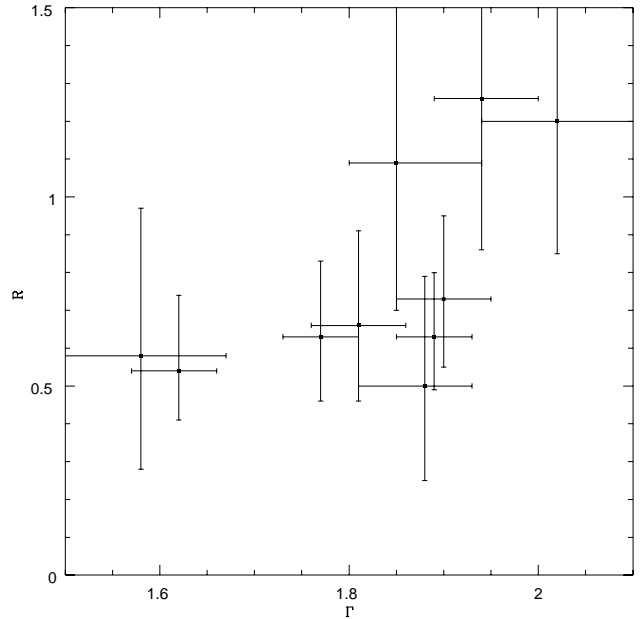}
\caption{$R$ plotted versus $\Gamma$.
}
\end{figure}

\begin{figure}
   \centering
   \includegraphics[width=9.cm]{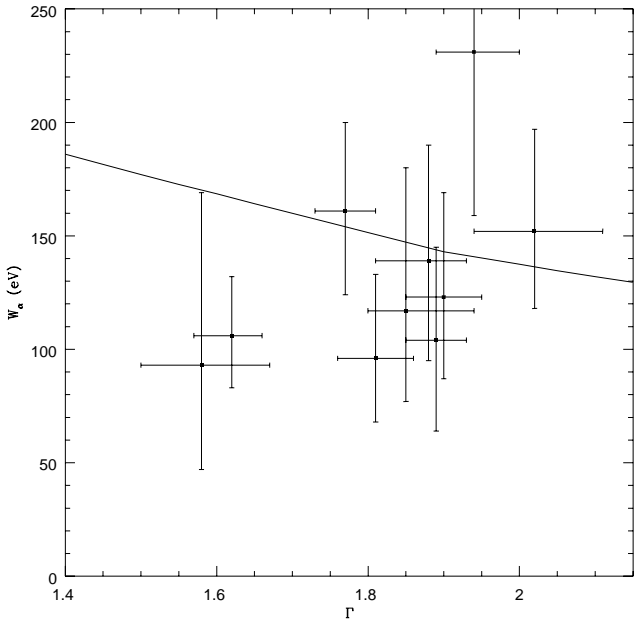}
\caption{$W_{\alpha}$ plotted versus $\Gamma$. The line, arbitrarily 
normalized,
describes the trend expected (derived from George \& Fabian 1991).
}
\end{figure}

\subsection{$R$ and $W_{\alpha}$ versus $\Gamma$}

 In a diagram of $R$ {\it versus} $\Gamma$, scatter is expected due 
to differences in orientation. Fig. 2 shows scatter
in $R$ only above $\Gamma$=1.8, where the largest values are confined
together with values comparable to those found below. While it could be
due to a statistical fluctuation in the sampling of the angles, 
this mere fact provides some evidence of a trend in 
$R$ to increase {\it on average} with increasing $\Gamma$. This 
is reflected in the Spearman rank--order correlation coefficient $r_{\rm s}$,
which is positive and equal to 0.67, with a significance level for $r_{\rm s}$ 
deviating from zero equal to 0.03. The marginal significance of this
evidence is further
weakened by the consideration (see Sect. 1) that the two variables $R$
and $\Gamma$ remain far from independent also in the fits to our very broad
band data, as can be appreciated with a simple exercise: by fitting the data
of each observation with $\Gamma$ fixed at the two extremes of its error bar, 
the values of $R$ obtained differ typically by as much as 60 to 80\% of their
90\% confidence intervals. In any case, the trend in our sample (admittedly much
smaller, 9 objects as opposed to 23) is definitely much
shallower than the correlation found by Zdziarski et al. (1999; see also Sect. 4.3
for a supplementary cause likely to contribute to the difference).

In addition to the one due to differences in the orientation, a scatter
due to differences in abundance is expected in a graph of $W_{\alpha}$ versus
$\Gamma$. A weak anticorrelation is predicted, with $W_{\alpha}$ decreasing
with increasing $\Gamma$, as mentioned above. The line (arbitrary 
normalization) describing this effect
(deduced from George \& Fabian 1991) is plotted in the
graph of $W_{\alpha}$ versus $\Gamma$ in Fig. 3. Notably $W_{\alpha}$ shows
instead a weak tendency to increase, on average, with $\Gamma$, which resembles 
that obtained by Lubi\'nski \& Zdziarski (2001, their Fig. 1).
 This is reflected in the sign of $r_{\rm s}$=0.57, which, contrary to the
prediction, is positive. The significance level for $r_{\rm s}$ deviating 
from zero is however 0.08, three times larger than in the previous case.
We note, though, that in the fits the variable $W_{\alpha}$ is less dependent
on $\Gamma$ than the variable $R$: the same exercise performed for $R$, and just mentioned,
yields differences in $W_{\alpha}$ which are at most, and  typically less than
35\% of their 90\% confidence interval. Thus we are inclined to regard the
evidence of a shallow trend in $W_{\alpha}$ increasing {\it on average} 
with $\Gamma$, albeit very marginal, no more so than it appears to be for $R$.

\subsection{$E_{\rm f}$ versus $\Gamma$}

The plot of $E_{\rm f}$ versus $\Gamma$ in Fig. 4 appears to confirm 
the existence of a correlation, with $E_{\rm f}$ increasing on 
average with $\Gamma$, noted first in BeppoSAX data by Piro (1999) 
 and then by Petrucci et al. (2001, their Fig. 6).
Although the variable $E_{\rm f}$ in the fits, very much like $R$,
is rather strongly dependent on $\Gamma$, and therefore one should not
disregard the possibility that our result might be biased by this effect, 
we note that in this case the correlation
coefficient $r_{\rm s}$ is equal to 0.88, and the significance
level for $r_{\rm s}$ deviating from zero is comfortably lower
and equal to 0.0007.
Discussing this correlation in terms of the mechanisms that give
rise to the main spectral component goes beyond the scope of this
paper (but see Petrucci et al. 2001). Our purpose here is to point 
out that, so long as one adopts the
empirical representation of the direct continuum as a power law with an 
exponential cut--off, the existence of this type of correlation could, 
if ignored, artificially contribute 
to a correlation between $R$ and
$\Gamma$: in fact, as {\it Ginga} data
do not allow to evaluate $E_{\rm f}$, by either neglecting the 
exponential cut--off
or keeping it at a fixed value irrespective of $\Gamma$ (Zdziarski et al. 
1999 choose $E_{\rm f}$ = 400 keV fixed), one would
systematically obtain lower values of $R$ for lower values of $\Gamma$,
for the reason already mentioned in Sect. 1. As an example of the interplay
between $E_{\rm f}$ and $R$,
we show the case of NGC 3783, whose $\Gamma$ has an intermediate value
in our sample: after excluding the cut--off we obtain $\chi^2$=212.1/149
($\Delta\chi^2$=61.7 demonstrates the high significance of the cut--off in
the fit given in Table 2) and a best fit value of $R$=0.38 as opposed
to $R$=0.63. Although the effect is sizeable, it alone is probably insufficient
to explain the difference in steepness of the correlation
found in the {\it Ginga} sample by Zdziarski et al. (1999) and in our sample,
but it does represent an explanation of the offset
between {\it Ginga} and BeppoSAX results noted, and tentatively attributed
to calibration differences, by the same authors.

\begin{figure}
   \centering
   \includegraphics[width=9.cm]{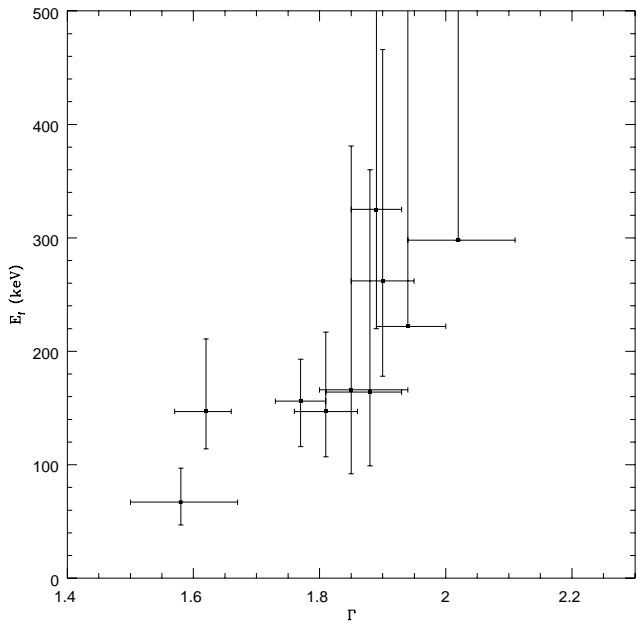}
\caption{The cut--off energy $E_{\rm f}$ plotted versus $\Gamma$.
}
\end{figure}

\section{Discussion and conclusions}

In confronting theory of reprocessing in AGN and observations, first
comes a test of the predicted correlation between the strength of
the reflection and that of the iron line fluorescence relative to
the primary photons in the PL, from material mainly al low ionization,
as indicated by the observed energy of the line. Our test, while 
improving on it, does confirm the result originally obtained by
Nandra \& Pounds (1994), who compared the mean sample values of $R$
and $W_{\alpha}$ with expectations from a face--on slab. We take
properly into account the dependence of $W_{\alpha}$ on $\Gamma$
in each observation and show that, be $R$ intrinsically equal to
one or otherwise, the amount of fluorescence is close to what
expected from the strength of the reflection, and that a 
modest variance in the iron abundance is likely present. In doing that,
we assume that the fairly substantial ``narrow'' contribution to
the line, which from {\it Chandra} and XMM-{\it Newton} observations appears
to be a rather common feature in this type of AGN, is associated
with optically thick gas. This assumption needs to be checked in
the future, at present, to our knowledge, it is observationally
confirmed in only one of the objects in our sample, NGC 5506.

We have then tested the existence of a correlation in $R$ and in
$W_{\alpha}$ with respect to $\Gamma$. In both cases we find a marginal
evidence of a very gradual increase, on average, with increasing spectral slope.
The trend in $W_{\alpha}$ is qualitatively similar to that 
found in the larger sample of ASCA observations (gaussian line fit) 
by Lubi\'nski \& Zdziarski
(2001, their Fig. 1), while the one that we find in $R$ is  
much shallower than
that found by Zdziarski et al. (1999) in {\it Ginga} data. Partly
this discrepancy could be due to the existence of a correlation,
of which we have found evidence in our sample,
between the e--folding energy
$E_{\rm f}$ (spanning from about 80 keV to more than 300 keV) and 
$\Gamma$, which we took automatically into account in our fits,
while Zdziarski et al. (1999) adopted a constant value equal to 400
keV. Thus our assessment, based on an admittedly much smaller sample
of objects, is that the same phenomenological description of the
spectral components, which are relevant to this issue, does not
confirm the existence of a strong correlation between the relative strength
of the Compton reflection, and the slope of the power law, while
it provides some evidence that a very gradual correlation,
and only on the average, for this quantity as well as for the iron line equivalent width,
cannot be altogether excluded, but need to be confirmed with a sample of objects
much larger than could be studied with BeppoSAX.

We conclude by commenting on three further limitations  
of this work.
The first has to do with the narrow component of the line. On this matter
further observational steps, which could not be pursued with
BeppoSAX alone, are required, to disentangle the broad (accretion
disk) component from the ``narrow'' component of the line,
and simultaneously the fraction of the reflection associated
with either of them. As recently shown by Malzac \& Petrucci (2002)
the nearly constant contribution by distant and thick matter, combined with
a power law spectrum pivoting, as it varies in intensity,
around energies in the range 1 to 10 keV, can introduce
correlations between the reprocessing parameters and 
$\Gamma$. Their attempt to explain in this scenario the paradoxical
result obtained on NGC 5548 by Chiang et al. (2000), mentioned in Section 3,
is particularly suggestive.

The second limitation concerns the ionization structure and radiative transfer
properties of the reprocessing material. Possibly with the exception
of a fraction of line emission in NGC 5506 and Mrk 509, in our sample
the adoption of a single zone, low ionization structure of the reflector
seems fully justified. However, Done \& Nayakshin (2001) have developed 
a disk model with a sharp transition from outer regions
of low ionization to inner regions with a thick skin, highly or
fully ionized, where the reflected continuum would be hard to 
distinguish from the direct continuum. In this model obviously the
reflected fraction pertaining to the outer regions would present
a value of $R$ substantially less than unity. Moreover, their
simulations of this model with the {\it Ginga} response, when fitted 
with the same model used by us, show that both $R$ and $W_{\alpha}$
turn out to correlate with $\Gamma$ in a rather gentle manner, qualitatively
similar to our results. It would be interesting to see this
exercise repeated with the wider band BeppoSAX response.

The third limitation has to do with the shape of the spectrum of
the primary component. The conventional exponential factor adopted
is useful to reproduce the steepening of the PL at high energies,
as required by the data, but clearly the value of $R$ which obtains
is very sensitive to this shape. When using instead models of
the continuum based on anisotropic comptonization in a disk corona
(Haardt \& Maraschi 1991, Haardt 1993) in fitting a sample of BeppoSAX 
observations, Petrucci et al. (2001) have shown that the value of $R$ 
turns out different, typically larger than that provided by a description
like the one used by us.

Thus on the whole there is still a very ample margin of progress
to be made, before the confrontation between theory and observations
could be accomplished in a more satisfactory way.

\begin{acknowledgements}
The BeppoSAX satellite is a joint Italian--Dutch program. We are grateful 
to the BeppoSAX Scientific Data Centre (now ASI--SDC) for assistance.
GCP thanks Fabio La Franca for help on a problem in statistics.
This work was partially supported by the Italian Space Agency (ASI), and by the
Ministry for University and Research (MIUR) under grant COFIN--00--02--36.
POP acknowledges a grant of the European Commission under contract ERBFMRX-CT98-0195. 
\end{acknowledgements}

\end{document}